# JET ALGORITHMS.
# WRAPPING UP THE SUBJECT

F. V. Tkachov

Institute for Nuclear Research of Russian Academy of Sciences, Moscow 117312, Russia
ftkachov@ms2.inr.ac.ru, http://www.inr.ac.ru/~ftkachov

This postscriptum to the theory of jet definition [hep-ph/9901444]
summarizes the points which did not find their way into the main text.

*Introduction*   1

I've just been told that the subject of jet definition can never be wrapped up.[1] What the title actually means is that this talk is supposed to close the series [1]–[5] in which I was presenting the theory of jet definition [6]–[9] as it evolved. The format and atmosphere of the QFTHEP workshops made them an ideal venue for that. In particular, all my useful interactions with experimentalists occurred or started at QFTHEPs — including the contact with physicists from D∅ which resulted in the dotting of an i in the observation of the top quark (see Sec. 4). So, first of all:

**Thanks to the organizers!**

*Theoretical synthesis of the optimal jet definition*   2

I do believe that the subject of jet definition is closed in the sense that every thing relevant and important has found its proper place in the theory of optimal jet definition (OJD):

- metaphysics (what is measurement? [7]);
- mathematical statistics (the theory of quasi-optimal observables whose relevance extends beyond jets to any parameter estimation problems where theoretical input is represented by a Monte Carlo event generator [11]);
- "high mathematics" (the surprising relevance of the so-called *-weak topologies for description of calorimetric detectors and hadronic energy flow [7]);
- quantum field theory (the central role of energy correlators and their fundamental expression in terms of the energy-momentum tensor [8]);
- QCD (the issues of IR safety, etc. [12]);
- numerical mathematics (algorithms for minimum search in 2K dimensions) and software engineering (the use of advanced software development tools [13]) culminating in the design of a fast and robust Optimal Jet Finder [5]; the code is available from [14]).

There is a remarkable synergy between different pieces, and none can be omitted without damaging the whole. From the above list, it should be clear why it took 25 years to solve the problem of jet definition.

*Principles of the theory*   3

The theory is systematically developed in two long papers [7], [9]. It is based on the following metaphysical realizations, treated in detail in [7]:

(I) **An absolute importance of the subject.** Indeed:

- data processing algorithms (=observables) are a meeting point of experimental data and theoretical formulae;
- hadronic jets are a way to observe quarks and gluons;
- jets are *everywhere* in HEP.

Realization of the absolute importance of the subject made me dismiss the sentiment[2] that any reasonable piece of Fortran code can, in principle, be accepted as a jet definition. The sentiment is only an expression of theoretical frustration at the 25 years' failure to find a solid foundation on which to build a systematic theory of jet definition.

(II) The theory of jet definition should be tightly linked to **quantum field theory**, *the* theory of fundamental interactions of which jets are a visible manifestation. In fact, my very first impression from the first explanations I received about jet algorithms (from W. Giele at Fermilab in 1992) was a surprise at how *un*-quantum-field-theoretical the popular jet algorithms were.[3] It was obvious to me as the discoverer of the algorithms on which the flourishing industry of NNLO QCD calculations is based ([16], [17]) that the conventional jet definitions do not conform to QFT at the basic kinematical/structural level, resulting in complications for systematic theoretical analyses.

---

[1] The person who said that announced after the talk his willingness to eat his words ☺

[2] Voiced by M. Mangano at [15].

[3] There is a paradox here because G. Sterman and S. Weinberg each published a textbook on QFT. I have an explanation for the paradox but it has nothing to do with the subject of jets per se.

(III) Jet algorithms must be fully understood as an instrument of **physical measurement.** An implication is that the analysis of measurement errors, including fluctuations due to the statistical nature of quantum theory predictions — most notably, how such errors survive the jet-algorithms-based data processing — must be a central issue to be addressed by the theory. This line of thought led me to investigate the following fundamental analogy:

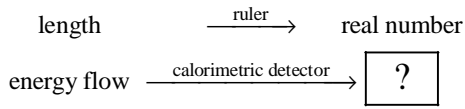

Filling the empty slot was the key issue I was preoccupied with in [7].

In the practical aspect, ref. [7] introduced:

(i) A large new class of *generalized shape observables* (the so-called *C*-algebra), some interpreted as non-integer jet numbers, others — so-called spectral discriminators — sensitive to the structure of multi-jet substates. The motivation was to illustrate that:
— jet algorithms are, in principle, not needed to extract the physical information usually obtained by their intermediacy;
— such information is best extracted using generalized shape observables (this became a theorem in my second paper [9]); and finally that
— jet algorithms are, from this viewpoint, only a means for approximate construction of such observables.

(ii) The notion of *regularization* of event cuts as an effective means to suppress statistical fluctuations induced in answers and thus enhance the signal/background. (Ref. [9] offered a more systematic treatment and provided new examples and prescriptions for regularizations.)

(iii) A binary recombination algorithm with an *optimal recombination criterion* which was derived uniquely from first principles (within the assumption of the binary recombination scheme) and which turned out to be a geometric mean of the JADE [18] and Geneva [19] criteria. (The optimal criterion was generalized in [9] into a global recombination scheme to become the optimal jet definition.)

In effect, the paper [7] introduced a novel, systematic point of view on jet algorithms as a means of construction of observables for precision measurements; it is, in a sense, purely kinematical in that it starts from the analysis of measurement errors and their evolution in data processing algorithms, so that on the surface, dynamical issues play a subordinate role (see, however, the discussion in Sec. 3.1).

After the paper [7] was finished, a well-meaning theorist advised me that I "should now do some data processing". The advice was ignored; the complexity of the problem called for some division of labor; I just did what I could in my situation, and went to great lengths[4] sparing no effort[5] to make my findings read by theorists and experimentalists alike.

As a reward for all the agony (not counting the damage my other projects suffered due to a few years' neglect), the paper [7] earned me exactly three external citations. To the experimental one I'll turn in Sec. 4. Of the other two, one was an incorrect attempt to literally implement an ancillary interpretation quoted in a footnote of [7] (see [9] for details). The other was a prominent but non-specific citation in a review talk [20].

Behind the scenes, there was (among other things) an earlier negative report on [6] which expressed a conviction that "kinematics" is not what the theory of jet algorithms is about — and that the QCD dynamics was the real issue. My reply (systematically presented in [7]) was that it may not be healthy to proceed to the dessert of dynamics without first doing away with the soup of setting up the kinematics correctly.

In fact, there are two potentially conflicting issues:

(a) Which jet definition is best for theoretical calculations?

(b) Which jet algorithm is best for extracting maximum information from experimental data?

The theory of the popular $k_T$ algorithm [21] simply misses (b) and, settling on an ad hoc algorithmic scheme (successive binary recombinations), focuses on (a). On top of that, convenience of theoretical calculations is judged from the standpoint of the Sudakov-Lipatov method of leading logarithmic approximation. No expert in diagrammatics can fail to appreciate the beauty of Sudakov's 1956 paper [22] — or to be frustrated (at least at first) by L.N. Lipatov's calculational tornado — yet to me, the LLA techniques is rather an antiquated piece; there are much more powerful ways to handle diagrams [23].

In short, the argumentation behind the $k_T$ algorithm left me unimpressed.

---

[4] like spending a 36-hours-long birthday passing customs, waiting for a delayed plane and flying non-stop across the whole of Europe, the Atlantic, and a good part of USA — all this heroism in order to drop the preprint [7] right where the action was.

[5] The outrageously un-academic — but extremely effective —graphic on the first page was employed only after testing its effect on a few unwary colleagues ☺

*But what about dynamics?*[6]     3.1

In the language of theory, "dynamics" is nothing more than qualitative properties of QCD matrix elements. The most basic such property is the collinear singularities with the probability density behaving as

$$\sim \theta^{-1} d\theta \,, \qquad 3.2$$

where $\theta$ is the collinear angle of the emitted parton.

The leading power behavior 3.2 is exactly what determines the specific form of the usual conditions of IR safety as explicitly formulated in [24]. So making one's observables IR safe already includes the most important dynamical information into the picture. This is the case e.g. with OJD — both the $2 \to 1$ recombination version of [7] and the global $N_{\text{part}} \to N_{\text{jets}}$ variant of [9].

If the collinear singularities were more severe, e.g. $\theta^{-2}$, then restrictions would have to be imposed not only on the values of observables but also on their first derivatives, and the entire theory [9] together with the resulting OJD would have changed accordingly.

Compared with the leading power behavior, logarithmic corrections — resummed or not — play a subordinate role (this is particularly clear in the context of the theory [7], [9] which offers a constructive model for the notion of "physical information of the event").

So as no surprise comes the evidence (see Sec. 3.3) that the behavior of various flavors of $k_T$ algorithm would differ from the binary version of OJD within the uncertainties of the $k_T$-type algorithms.

Moreover, ref. [9] contains an argument which demonstrates that OJD is optimal not only in the kinematical sense but — which was not a priori expected — also in a properly formulated dynamical sense. This shows that the IR safety (reflecting the singular structure 3.2) is all that really must be taken into account in the definition of jets; anything on top of that is best treated as something only theorists should worry about in their calculations. This conclusion is corroborated by the findings of [25] to which I now turn.

*Comparison of different binary recombination algorithms*     3.3

Another theoretical paper which came to my attention after the completion of [9] is the comparison of various versions of the iterative $2 \to 1$ recombination scheme performed by acknowledged experts in such matters [25]. The comparison contains a number of results which are visible confirmations of some a priori claims made in [7] and [9]; for many, such explicit calculations may be more convincing than the logical arguments.[7]

For convenience of discussion I divide the algorithms considered in [25] into two groups: the "good" algorithms (variations of the venerable Luclus[8] and the $k_T$) and the "bad" algorithms (JADE and Geneva).

It was, of course, superfluous to include into comparison the optimal criterion of [7] because its behavior can be easily deduced from the results for the two "bad" algorithms. The latter fact being not mentioned in [25], here are some comments.

The optimal, Geneva, and JADE criteria have, respectively, the following forms:

$$(E_a + E_b)^{-1} E_a E_b \times (1 - \cos\theta_{ab}) < y_{\text{cut}} \,, \qquad 3.4$$

$$(E_a + E_b)^{-2} E_a E_b \times (1 - \cos\theta_{ab}) < y_{\text{cut}} \,, \qquad 3.5$$

$$E_a E_b \times (1 - \cos\theta_{ab}) < y_{\text{cut}} \,. \qquad 3.6$$

The optimal criterion happens to be a geometric mean of the other two. It is perfectly clear that its behavior should also be a median one.

Now the several graphs presented in [25] (e.g. Figs. 18, 19, etc.) show that the behavior of JADE is always the worst, Geneva behaves erratically, and the "good" algorithms populate the band between JADE and Geneva.

The band turns out to be pretty wide (proving that the effect of what one expects to be insignificant algorithmic variations such as changes in the order of recombinations, is large and not incomparable with the difference between the "good" and "bad" algorithms).

It is also perfectly obvious that the optimal criterion is bound to be among the "good" algorithms. An implication is that its global analog — the OJD — must behave similarly to the "good" algorithms.

Now on to the experimental citation of [7].

---

[6] A question by S. Catani at [15].

[7] A difficulty with synthetic solutions is that they often involve logical patterns which abstract experiences from other problem domains (where such experiences may be common knowledge) rather then the specific one being dealt with (where such principles can be a novelty and, as such, are greeted with suspicion).

[8] Note that the definition of Luclus changed compared with the 1983 original [26]: the 1998 Luclus [25] is equivalent to the 1995 optimal criterion [7] with the angular factor square-rooted.

Recall that the two collaborations which reported the discovery of the top at FNAL in 1995, CDF and D0, were in very different positions: CDF had the SVX *b*-tagging hardware which proved to be a powerful means for selecting the events with top. In particular, at the time of the top discovery in 1995, D0 could not see the top signal in the all-jets channel. So there was a strong competitive pressure on D0 to seek alternative methods. The preprint release of [7] could not be more timely.

I'd like to emphasize that the initial contact with D0 physicists occurred at QFTHEP'93 [1]. Another peculiar circumstance is that both physicists (A. Klatchko and D. Stewart) who are to be credited for bringing the new theory to the attention of the D0 all-jets team, had to leave physics.

At QFTHEP'93 A. Klatchko picked up a copy of my talk and brought it to Fermilab. The copy was inherited by D. Stewart. Although about to leave physics too, he made the effort of finding me at Fermilab (where I was able to visit thanks to a G. Soros grant and the hospitality of FNAL theorists), showed me around the D0 detector and brought me into contact with other members of the all-jets group (S. Ahn and H. Prosper), and we had a nice long discussion about regularization of cuts and spectral discriminators. N. Sotnikova (a member of both D0 and the QFTHEP team) ensured that the draft of [7] was well advertised and plentifully available at D0. The next time I heard about the D0 all-jets group was at QFTHEP'96 from E. Boos who mentioned some striking results but could not provide any details. I tried to get into contact with D0 but failed. I heard nothing from them since then — and switched to asymptotic expansions of Feynman diagrams, the subject neglected since 1992 in favor of jets — until I ran across ref. [10] while at CERN in March, 2000. Their page 59 contains the following passage which I cannot resist quoting:

"… Additional cuts are applied to remove events with noise from the Main Ring. After these cuts, about 280,000 events remain. At this stage, the signal to background ratio is about 1:1000. Requiring an SLT *b*-tag increases the signal/background by about an order of magnitude, leaving 6000 events.

To make further progress, D0 performs a multivariate analysis using thirteen variables, described briefly in Table 10. The principal ones are $H_T$, $H_{T3} \equiv H_T - E_T^{\text{jet1}} - E_T^{\text{jet2}}$, the average jet count $N_{\text{jets}}^A$, the aplanarity $\mathcal{A}$, the centrality $C \equiv H_T / \sum_{\text{jets}} E^{\text{jet}}$, and the transverse momentum of the muon $p_T^\mu$. A particularly powerful (and unusual) variable is the average jet count, defined by

$$N_{\text{jets}}^A = \int_{15\,\text{GeV}}^{55\,\text{GeV}} E_T N(>E_T)\, dE_T \bigg/ \int_{15\,\text{GeV}}^{55\,\text{GeV}} E_T\, dE_T, \tag{17}$$

where $N(>E_T)$ is the number of jets with transverse energy greater than $E_T$. This variable, inspired by the work of Tkachov[90] is interesting in that it assigns a nonintegral "number of jets" to the event.

These thirteen variables are combined using feed-forward neural networks …" ([90] = [7])

The final result of the procedure is observation of the excess of events containing top over background in the total cross section at the level of 3.2$\sigma$.

To appreciate the work behind the observable (17), note that it has no direct analog in [7]. I can think of the following way to arrive at it. In [7], the so-called spectral discriminators were introduced which, for narrow jets, allowed representations

$$\int d\sigma\, w(\sigma) \sum_j \delta(\sigma - S_j), \qquad 4.1$$

where summation runs over all (multi-)jets of the event and $S_j$ is a variable like invariant mass (or transverse energy) of the (multi-)jet. This kind of observables are not usually considered, and D0 must have simply ignored what I wrote about how such observables can be defined bypassing jets, and focused on the simplest case, i.e. distribution of event's jets along the axis $E_T$, and tried to see how the resulting patterns for background events and top events differ. They might have replaced $\delta$'s with the equivalent integral quantities $N(>E_T)$ (similarly to how this is done in the probability theory to avoid directly dealing with singular probability densities) and then used event generators to study the corresponding patterns for background events and events with top to arrive at their observable (see Sec. 4.2).

At this point, I cannot help reminiscing how a prominent QCD expert told me he'd found ref. [7] incomprehensible — apparently, because there were in it no leading logarithms, no parton distributions — in short, none of what QCD theorists swear by. Another expert advised me that the things I discussed were not what experimentalists really needed. Also, shortly after the release of [7] an American theorist P. ventured to run tests of spectral discriminators. Despite obtaining pictures which fully agreed with the qualitative expectations of [7], P. strangely made an exactly opposite conclusion. My guess is, he simply coded the formulae of [7] as is without trying to understand their meaning, and naively expected to see exactly the bumps I draw there. Instead, he saw a shoulder, and a steeply rising background to the left of it. The presence of the shoulder was a clear indication that the signal was there, and one only had to study its evolution and to figure out a simple way to extract the numerical information (e.g. a wavelet filter would do nicely). But, apparently, there are easier ways for a theorist equipped with a computer and a MC generator to produce a publication than deep experimentation with novel ways to process data. (For fairness' sake, the dominant software engineering platform in HEP — Fortran, C, etc. — does not encourage such experimentation.)

### *Power of regularizations* 4.2

An important trick in the construction of good observables is ***regularization***. This is a concept widely used in applied mathematics (see [7] for references). In [7] and [9], I only gave simple examples to illustrate potential benefits of regularizations (which, by the way, come in many flavors, although with the same underlying principles). The above D0̸ observable (17) offers a great real-world discovery-class example of how useful a regularization can be.

Indeed, suppose one has noticed that top events tend to produce more energetic jets. Then one would consider the observable $N(>c)$ ("the event has $N$ jets with energies above $c$") with some $c$. An integer-valued observable is discontinuous, and its discontinuities are, on general grounds, sources of non-optimality (loss of physical information due to unnecessary enhancement of statistical fluctuations in the transition from raw data to the values of the observable; see [9]). Such non-optimalities are eliminated by introducing a continuous regularization, e.g. in the form of a smearing over the cut $c$. General principles do not fix details like the weight used for the smearing. At this point some experimentation is needed. Finding an optimal shape of regularization, however, is a much more specific task than the original one.

Note that the smearing as a way to regulate a hard cut was not explicitly mentioned in [7] (although smearings were mentioned in other contexts).

A conclusion from the D0̸ experience is as follows:

> Learn to use regularizations.

### *Optimal jet definition: the final form* 5

The second part of the theory, ref. [9], improved upon the first in almost every point. I now had a *proof* that observables which yield the theoretically best precision for a particular parameter are always the generalized shape observables; jets-based observables can only yield approximations to such optimal observables. As a result:

> The entire theory of jet algorithms now reduces to studying ways of construction of such approximations.

It is this embedding of the problem of jet definition into a broader context which provides an unambiguous criterion for deciding which algorithm is best:

> The best jet algorithm is that which allows one to construct the best observable for measurements of a given parameter; the best observable is that which yields the best precision for the parameter.

Second, I now derived the most general $N_{\text{part}} \to N_{\text{jets}}$ recombination criterion for finding jets. The criterion required minimization of a multi-axes generalization of the well-known thrust. (An e-mail from W. Giele about a forthcoming session of the jet definition working group at FNAL in March, 2000 stimulated exploration of algorithmic implementations of OJD, resulting in an efficient code [5].)

Third, there were systematically motivated ways to treat the problem of non-uniqueness of jet configurations via association of multiple (appropriately weighted) jet configurations with the same event. For details, the reader is referred to [9].

### *Quasi-optimal observables* 5.1

Pondering the D0̸ observable (17), one is faced with a natural question: Is there a way to obtain such "particularly powerful" observables in this and similar problems with less guesswork than had been involved in the finding of (17)? The answer is yes, and the corresponding prescriptions are called ***the theory of quasi-optimal observables***. In view of its general importance, it was posted separately [11] rephrased in the jargon of parameter estimation of mathematical statistics. The theory is simple and closely related to the Rao-Cramer inequality; see [9] or [11].

Quasi-optimal observables are a means to approach the quality of the maximal likelihood method in situations where its application is problematic such as encountered in HEP (infinite dimensionality of the underlying event space and a probability density in the form of event generator rather than an explicit formula). Although maximal likelihood is widely used (see e.g. [10]), the method of quasi-optimal observables is simpler, more flexible, and offers new options both for design of general-purpose software tools and for data processing in specific applications.

For instance, instead of the observable (17), I would simply MC-generate a quasi-optimal observable in the 6-dimensional space of the variables $E_T^{\text{jet}}$ (e.g. using a rectangular grid; the formal parameter to be estimated would be the coupling to produce top-antitop pairs). If constructed with enough care, such an observable might well obviate the need for neural networks.

(This would certainly be the case for a quasi-optimal observable constructed for the variables including the six $E_T^{\text{jet}}$ plus all the variables from Table 10 which cannot be expressed in terms of these.)

> ➡ It would be interesting to develop universal adaptive software to automatically generate quasi-optimal observables given a sample of events and a number of (more or less ad hoc) variables.

(It looks like this could be done without recurring to fancy techniques such as neural networks etc.)

A practical conclusion is this:

> Learn to construct quasi-optimal observables.

Unless I missed something in [10], the top mass has not yet been measured in the all-jets channel at D0. This could be an excellent starting point.

### *The global $N_{\text{part}} \to N_{\text{jets}}$ recombination algorithm*  5.2

The derivation, description and discussion can be found in [9]. A convenient brief summary of the definition is given in [5]. Suffice it to say the following:

- OJD prescribes to determine the jet configuration via minimization of shape observables which generalize the venerable thrust to *n* axes.

- OJD is similar to a cone algorithm[9] in that there is a parameter, *R*, which limits the cone radius, and the resulting cones are less irregular than with binary recombinations.

- Being a global version of the binary recombination algorithm of [7], its behavior and properties (e.g. stability) can only be better than for the binary recombination version (which as we saw in Sec. 3.3 should be on a par with other "good" jet algorithms[10]).

- The speed of jet search is expected to be $O(N_{\text{part}} \times N_{\text{jets}})$; compared with the $O(N_{\text{part}}^2)$ behavior of binary recombination schemes (including $k_T$), this property alone might make OJD the jet algorithm of choice for future colliders such as LHC where multiplicities are much higher.

- OJD is intrinsically connected with the theory of quasi-optimal observables. As a result, OJD (unlike binary recombination algorithms, whether optimal or $k_T$) offers natural regularizations to solve such problems as the potential non-uniqueness of jet configurations (which reflects the fact that different parton configurations may generate the same hadronic final state).

### *On the difficulty of computer implementation*  5.3

From [25] I learned that generalizations of thrust were proposed for jet definition in the early 80's [28]. The exact form of OJD (in particular, the treatment of soft energy via an additive term) could, of course, not have been guessed a priori. On the other hand, ref. [25] concluded that jet search based on optimization of thrust-like shape observables was computationally prohibitive. To this I note the following:

The code and data structures of the OJD implementation described in [5] take under 300K of RAM for LEP2 events, and the algorithm finds jets for such an event in a fraction of a second on a modest Pentium.[11] Such an algorithm could be implemented even on a 1984 PC with 640K of RAM.

What was lacking was not hardware or even software (the old Fortran IV not to say Turbo Pascal 1.0 would have been sufficient) but expertise in numerical mathematics and software engineering as well as determination. Determination could have only come from realization of the theoretical depths behind OJD. The software engineering aspect is discussed in the notes collected in [29]. Here I would only like to point out that the displacement of Fortran in favor of C++ (rather than strictly type-safe languages such as any version of Oberon-2; see e.g. [13]) in the physicists' software engineering is a ***disaster of historic proportions*** (and a good part of the explanation of inability of theorists to accomplish complete 1-loop calculations for LEP2). It is also absolutely irrational in view of the existence of simple and powerful alternative in the form of the Oberon-2 family of languages representing the brilliant tradition of design of N. Wirth and his school starting from the old Pascal (see [29] for more on this).

> The harnessing of the digital revolution being a major infrastructure problem in physics as elsewhere, a competent and responsible policy of steering the HEP community towards the use of safe programming languages — and redirecting human energy released thereby into more productive activities than debugging C++ codes — would have a major positive impact on physics.

---

[9] I incorrectly stated in a talk that OJD is *equivalent* to a cone algorithm for events with well defined jets. However, the cone radius in OJD depends on the distribution of energy around the jet axis, so that the techniques of energy corrections developed for fixed cone algorithms would have to be modified.

[10] This has been confirmed in a number of un-premeditated tests run on realistic MC event samples by Pablo Achard [27]. In at least one graph I saw with my own eyes, OJD resolved partons significantly better (with an efficiency close to 100%) than $k_T$ (~70%) in the entire interval of rapidities where both algorithms yielded meaningful results.

[11] For the events tested by Pablo Achard [27], the OJD implementation of [5] was somewhat slower — but not significantly — than $k_T$. Note that ref. [5] focused on feasibility studies and did not attempt to implement all possible algorithmic optimizations.

## 6  Conclusions

To the best of my understanding, the subject of jet definition as a theoretical problem is closed.

The most important next task in regard of jet-related data processing appears to be for physicists to learn to use OJD for construction of quasi-optimal observables for specific applications.

Psychologically, the main difficulty may be to stop wasting human effort (and other valuable resources) on the obsolete jet definitions.

*Acknowledgments*

This work was supported in part by the Russian Foundation for Basic Research (quantum field theory section[12]) under grant 99-02-18365.[13]

I am grateful to V. von Schlippe for a comment, and to S. Karshenboim for suggesting a term.

Finally, it is my greatest pleasure to thank the organizers of the QFTHEP workshops (E.E. Boos, B.B. Levtchenko, V.I. Savrin and N.A. Sveshnikov ) for their invariable support over all these years.

---

[12] There is far too much to my taste of this sort of things that goes unrecorded, so here is my contribution: in 1998 when the above FNAL report was released, the HEP section of the Russian Foundation for Basic Research denied me support for my projects. If the support were not resumed by the quantum field theory section (and no, jets were not even mentioned in my grant application) you would most certainly not learn about OJD.

This must have something to do (as I realize post factum) with the old story of *Landau vs. Bogolyubov* reflected within the Russian theoretical physics in the form of a certain *gruppovschina*, with the two sections of RFBR influenced by the respective theoretical schools. I do not see what I personally have to do with that old story except a sufficient mathematical background to have drifted in my years of innocence to Bogolyubov's cathedra (I do own a complete set of the Landau and Lifshits textbooks too).

For fairness' sake, the fundamental reason for that bitter controversy lied not so much in particular personalities (the mistake usually made) but in the overcentralization of everything in Russia, so that the entire hierarchy is affected by idiosyncrasies of the individuals at the top multiplied by the centuries-old tradition of servility of subordinates.

There is an objective tendency for similar effects to occur wherever individuals are endowed with considerable administrative powers (Germany comes to mind, with perhaps servility replaced by discipline) and wherever resource-allocation politics finds fertile soil in complex hierarchies within which there often are no efficient checks to deter individuals from irresponsibility in spreading their "necessarily somewhat subjective" opinions.

[13] I wish I could benefit from even a tiny fraction of the funds used up by the two jet definition working groups (FNAL and Des Houches) which operated during 1999 while the work on the final version of [9] and on the software implementation of OJD [5], was in progress: a mere 0.1% would have been enough for a mighty upgrade of my computers. It is rather a paradox given the state of physics in Russia that up to this point, the work on the theory of jets and OJD was performed as a kind of charity for the international HEP community.

## 6.1  Appendix. The benefits of OJD compared with $k_T$

All the QCD argumentation behind $k_T$ boils down in the end to the proposition that with it, better theoretical calculations are possible. However:

• Experimental data are more valuable than theoretical numbers ⇒ Preserving information from data is more important than making it easier for theorists to do their calculations. OJD is optimal precisely because it preserves the experimental information in the best manner possible.

• As regards calculations: nothing compares with shape observables such as thrust in regard of calculability (LLA or NLLA) and amenability to theoretical analyses (power corrections), and OJD is based on such observables. Furthermore, there are better ways to do complex calculations than the archaic Lipatov-Sudakov techniques, namely, the systematic machinery of asymptotic operation [23] whose euclidean version, remember, is behind most of the well-known NNLO calculations in QCD.

• Further, the binary recombination scheme per se is flawed (cf. the large variations between different variants of "good" recombination criteria [25]).

• The $k_T$ algorithm behaves quadratically in the number of particles, and OJD may run faster for LHC multiplicities.

• OJD offers new options for a systematic construction of better observables — options not available within any conventional jet definition schemes.